\documentclass[aps,pre,showpacs,amsmath,amssymb,preprint]{revtex4}
\usepackage{bm}
\usepackage{graphics}

\begin{document}
\draft

\newcommand{\uu}[1]{\underline{#1}}
\newcommand{\pp}[1]{\phantom{#1}}
\newcommand{\be}{\begin{eqnarray}}
\newcommand{\ee}{\end{eqnarray}}
\newcommand{\ve}{\varepsilon}
\newcommand{\vs}{\varsigma}
\newcommand{\Tr}{{\,\rm Tr\,}}
\newcommand{\pol}{{\textstyle\frac{1}{2}}}
\newcommand{\ba}{\begin{array}}
\newcommand{\ea}{\end{array}}
\newcommand{\bea}{\begin{eqnarray}}
\newcommand{\eea}{\end{eqnarray}}
\title{
Abstract DNA-type systems
}
\author{Diederik Aerts$^1$ and Marek Czachor$^{1,2}$}

\address{
$^1$ Centrum Leo Apostel (CLEA) and Foundations of the Exact Sciences (FUND)\\
Vrije Universiteit Brussel, 1050 Brussels, Belgium\\
$^2$ Katedra Fizyki Teoretycznej i Metod Matematycznych\\
Politechnika Gda\'nska, 80-952 Gda\'nsk, Poland}

\begin{abstract}
An abstract DNA-type system is defined by a set of nonlinear kinetic equations with polynomial nonlinearities that admit soliton solutions associated with helical geometry. The set of equations allows for two different Lax representations:  A von Neumann form and a Darboux-covariant Lax pair. We explain why non-Kolmogorovian probability models occurring in soliton kinetics are naturally associated with chemical reactions. The most general known characterization of soliton kinetic equations is given and a class of explicit soliton solutions is discussed. Switching between open and closed states is a generic behaviour of the helices. The effect does not crucially depend on the order of nonlinearity (i.e. types of reactions), a fact that may explain why simplified models possess properties occuring in realistic systems. We explain also why fluctuations based on Darboux transformations will not destroy the dynamics but only switch between a finite number of helical structures. 
\end{abstract}
\maketitle

\section{Introduction}

There are certain aspects of the DNA dynamics that go far beyond standard fields of interest of genetic engineering. A striking example is the observation that DNA is essentially a universal Turing machine \cite{BL,Adleman}. The hardware of the machine consists of two strands (input-output tapes) and an enzyme plays a role of finite control. As opposed to our PC's that cannot work without electricity, the DNA machine is powered by networks of autocatalytic reactions. The double-helix structure is a trademark of the machine.

A new DNA helix is formed by a polymerase that moves along an existing strand and gradually, step by step, produces the complementary strand. This seems to be a continuous process that resembles the motion of finite control in a real-life Turing machine. The discrete time-steps of an abstract Turing machine can be obtained if one looks at snapshots of the evolution taken at times $t_n=n\Delta t$, say. Finally, the two helices start to unwind, and the two strands behave as the tapes of two new independent Turing machines. This process is called replication. 

The problem how to theoretically model the DNA dynamics is an old one. Between various notable theoretical constructions one should mention classical Hamiltonian systems 
(typically of a lattice type) \cite{Englanger,Yak,Yak1} or replicator networks 
\cite{Eigen,Schuster,HS}. 

The above approaches are well motivated. There is little doubt that DNA is a nonlinear dynamical system. It is also clear that the DNA evolution is a sequence of chemical reactions and we know how to translate kinetic schemes into differential equations with polynomial nonlinearities. Therefore no matter what networks of reactions one will invent, the result is in general a set of nonlinear equations for certain nonnegative variables. As opposed to the mechanical models, the kinetic equations are not directly linked to soliton systems.

The DNA dynamics does not look chaotic. One could even argue that organisms whose DNA evolves in a chaotic way would be eliminated by natural selection. Systems that are nonlinear and non-chaotic are almost certainly solitonic. Solitons are localized, propagate in a stable way, and are very resistant to external noise. Soliton equations naturally lead to 
`switches', that is, solutions of a kink type. This is why, the line of research that deals with Hamiltonian systems is concentrated on `DNA solitons'. Another argument in favour of DNA solitons is that soliton scattering can serve as a model of computing \cite{SQ1,SQ2,SQ3}. The DNA computer might be a soliton machine.

The latter observation points into another general class of dynamical systems: Those whose discretized dynamics (in the sense of $t_n=n\Delta t$ snapshots) is equivalent to some algorithm. The first example of such a link is the Toda lattice interpreted in terms of spectrum generating algorithms \cite{Symes}. Later on, the construction was generalized to a general class of `Lax equations' \cite{Przybylska1,Przybylska2}. The term `Lax equation' is often applied to equations whose right-hand side is given by a commutator. However, in the soliton case we need more than just a commutator: We need a Darboux covariant Lax pair 
\cite{Salle} and this is not exactly synonymous to a commutator. Equations with commutator right-hand sides may not be Darboux covariant. Or, even if they are Darboux covariant, the Darboux transformation can be inconsistent with some important constraints. 

To sum up, it looks like it would be ideal to work on DNA dynamics with soliton kinetic equations that have some sort of Darboux-covariant Lax representation. Such equations do, in fact, exist and can be characterized in quite general terms. To find them we are led by still another hint: The ubiquity of helical structures associated with DNA dynamics. The fact that the helices are formed (and deformed) as a result of the dynamics suggests their 3D parametrization of the form 
$\big(t,x(t),y(t)\big)$, where $t$ is time (i.e. position of the read/write head) and $x(t)$, $y(t)$ are related to kinetic variables. 
The kinetic variables must therefore occur in {\it pairs\/}. Alternatively, having an even number of dynamical variables one can always work with their complex representation $\xi(t)=x(t)+iy(t)$. 

Let us now try to derive a simple kinetic scheme that leads to a helix. At this stage we are satisfied by the `boring' case of a helix that does not replicate. Consider the following catalytic network \cite{SKoter}
\be
B &\stackrel{k}{\to}& A +B \label{B->A}\\
A+X &\stackrel{k'}{\to}& X +{(\rm products)}\label{A+X}\\
A+B &\stackrel{k}{\to}& A +{(\rm products)}\label{A+B}\\
Y &\stackrel{k''}{\to}& B \label{X->B}
\ee
Assume the reactions (\ref{A+X}), (\ref{A+B}) are 0-th order in $A$ and $B$, respectively. Then
\be
{[\dot A]}
=
k[B]-k'[X],\quad
{[\dot B]}
=
k''[Y]-k[A],\quad
{[\dot X]}=[\dot Y]=0.
\ee
If we define two new (in general non-positive) variables $x=[A]-(k''/k)[Y]$, 
$y=[B]-(k'/k)[X]$, we can rewrite the system as
$
i\dot \xi=k\xi
$
where $\xi=x+iy$. The system evolves as a harmonic oscillator. The curve $t\mapsto 
(t,[A(t)],[B(t)])$ is a helix. If
\be
H
=
\left(
\begin{array}{cc}
k & 0\\
0 & 0
\end{array}
\right),
\quad
\rho
=
\left(
\begin{array}{cc}
a & \xi\\
\bar\xi & b
\end{array}
\right),
\ee
$a\geq 0$, $b\geq 0$, $a+b>0$, $\dot a=\dot b=0$, then 
$
i\dot\rho=[H,\rho]\label{ivN}.
$
This is what we call a Lax form of the kinetic equations. 
If eigenvalues of $\rho$ are nonnegative at some $t=t_0$ then for any projector $P_A$ we have $[A]=\Tr\rho P_A\geq 0$. Let now $P_A$ and $P_B$ project, respectively, on 
$
\left(
\begin{array}{c}
1\\
1
\end{array}
\right)
$ and
$
\left(
\begin{array}{c}
1\\
-i
\end{array}
\right)
$.
Then $x=[A]-(a+b)/2$ and $y=[B]-(a+b)/2$, i.e. $(k'/k)[X]=(k''/k)[Y]=(a+b)/2$. 
The kinetic variables satisfy $[A(t)]=\Tr\rho(t) P_A$ 
and $[B(t)]=\Tr\rho(t) P_B$. The condition $[B(t)]>[A(t)]$ can be always satisfied by an appropriate shift of origin of the $(x,y)$-plane. What is interesting, the projectors $P_A$ and $P_B$ do not commute (projectors in a plane commute if and only if they project on either parallel or perpendicular directions, which is not the case here). This type of noncommutativity is an indication of a non-Kolmogorovian probabilistic structure behind the kinetics.

The above analysis explains the motivation for the formalism we develop in this paper. We will work with kinetic equations that possess a two-fold Lax structure. First of all, all these equations can be written as von Neumann type nonlinear systems. Secondly, each of these equations can be regarded as a compatibility condition for a Darboux-covariant Lax pair. 

The Darboux transformation will also play a two-fold role in our formalism. It will allow us to find nontrivial exact solutions of systems of coupled nonlinear kinetic equations. The transformation is known to switch between different solutions of the equation in a way that respects certain conservation laws (Casimir invariants). It is therefore also a natural formal object representing fluctuations along a symplectic leaf of the dynamics. This effect will lead to both mutation and formation of an open state in the associated helical structures. 

That nonlinear von Neumann equations form a dynamical framework for various traditionally non-quantum domains was the main message we tried to convey in 
\cite{PRE03}. In the present paper we lay special emphasis on the aspect that was implicitly present in all the examples we gave, but which at a first glance may not be completely obvious: The fact that von Neumann equations are not {\it analogous\/} but, in fact,  {\it exactly equivalent\/} to systems of kinetic equations. Any von Neumann equation is just a form of a Lax representation of a set of kinetic equations (this observation was implicitly present also in the analysis of replicator equations given in \cite{Pryk}, although technically this argument was not equivalent to the one we give below). This is why any integrable dynamical system that involves kinetic variables is suspected of being implicitly von Neumannian. 

In spite of its simplicity, the above statement is surprisingly difficult to digest by the community of chemists or molecular biologists. The roots of the difficulty seem to be related to the misleading term `quantum probability' suggesting that non-commutative propositions are restricted only to microscopic systems. 

Therefore, one element we will need to understand before launching on technicalities of soliton kinetic equations, is the structure of probability models that are behind soliton kinetic evolutions. We have seen already at the simple example discussed above, that the `concentrations' $[A]$ and 
$[B]$ were linked to noncommuting projectors $P_A$ and $P_B$. This type of noncommutativity is a typical feature of soliton kinetics and indicates that the probability calculus is here nonclassical. We therefore need to clarify some formal aspects of non-Kolmogorovian probability, and explain why this is the type of formalism that {\it necessarily\/} occurs in chemical kinetics. 

\section{Hilbertian model of probability}

A proposition is represented by a self-adjoint operator with eigenvalues 0 and 1, i.e. a projector $P=P^{\dag}=P^2$ acting in a Hilbert space $\cal H$. Probability is computed by means of the formula 
\be
p=\Tr P\rho\label{p}
\ee
where $\rho$ is a `density operator' (`density matrix') acting also in $\cal H$ and satisfying $\rho=\rho^{\dag}$ (since then probabilities are real: $\Tr P\rho=\overline{\Tr P\rho}=\Tr P\rho^{\dag}$), 
$\rho> 0$ (i.e. the eigenvalues of $\rho$ are non-negative) since then $p\geq 0$. The normalization of probability leads to  
$\Tr\rho=1$: If the set $\{P_j\}$ of orthogonal projectors is {\it complete\/} in the sense that $\sum_j P_j=\bm 1$ then
\be
\sum_j p_j=1=\sum_j\Tr P_j\rho=\Tr\rho.\label{12}
\ee
In a given Hilbert space there exists an infinite number of such complete sets. If a given problem can be represented entirely in terms of probabilities associated with a single complete set, then all the probabilities will satisfy criteria typical of Kolmogorovian models. 

If the kinetics is implicitly Hilbertian then all the dynamical variables, say $p_1$,
$p_2$, $\dots$, $p_N$, have to allow for a representation of the form (\ref{p}). If a given set of kinetic equations couples probabilities $p_1,\dots,p_N$ that do not belong to the same complete set then 
$\sum_{j=1}^N p_j\neq 1$. This happens whenever the set $\{P_j\}$ involves projectors that do not commute with one another since then necessarily $\sum_j P_j\neq\bm 1$. Actually, this is what we found in the simple kinetic model of harmonic oscillators discussed in the Introduction, where $P_A$ and $P_B$ did not commute and, hence, did not belong to the same maximal set. 

It follows that the conditions we impose on the kinetic equations can be reduced to the requirements imposed on the dynamics of the density operator $\rho(t)$: For all times 
$t_1<t<t_2$ in the interval of interest we must find (i) $\rho(t)=\rho(t)^{\dag}$, 
(ii) $\rho(t)>0$, and $\Tr\rho(t)={\rm const}<\infty$. 

A basis in the Hilbert space is given by a set 
$\{|n\rangle, \,\langle n|m\rangle=\delta_{nm}\}$. 
Matrix elements of the density operator are in general complex 
\be
\rho_{nm}&=&\langle n|\rho|m\rangle=x_{nm}+iy_{nm}\\
\overline{\rho_{nm}}&=&x_{nm}-iy_{nm}=x_{mn}+iy_{mn}=\rho_{mn}
\ee
and thus $x_{mn}=x_{nm}$, $y_{mn}=-y_{nm}$, $y_{nn}=0$. 

The diagonal elements $\rho_{nn}=x_{nn}$ are themselves probabilities since 
\be
\rho_{nn}&=&\langle n|\rho|n\rangle=\Tr P_n\rho=p_n=x_{nn}
\ee
where $P_n=|n\rangle\langle n|$. Now, let 
\be
|jk\rangle = \frac{1}{\sqrt{2}}\big(|j\rangle+|k\rangle\big),\quad
|jk'\rangle = \frac{1}{\sqrt{2}}\big(|j\rangle-i|k\rangle\big), 
\ee
$P_{jk}=|jk\rangle\langle jk|$, $P'_{jk}=|jk'\rangle\langle jk'|$. Then
\be
x_{jk}
&=&
p_{jk}-\frac{1}{2}p_{j}-\frac{1}{2}p_{k},\\
y_{jk}
&=&
p'_{jk}-\frac{1}{2}p_{j}-\frac{1}{2}p_{k},\\
\rho_{jk}
&=&
p_{jk}+ip'_{jk}-\frac{e^{i\pi/4}}{\sqrt{2}}(p_j+p_k),\label{p+ip'}
\ee
where $p_{jk}=\Tr P_{jk} \rho$, $p'_{jk}=\Tr P'_{jk} \rho$.
It follows that a single density operator $\rho$ encodes three families of probabilities: 
$\{p_n\}$, $\{p_{jk}\}$, and $\{p'_{jk}\}$. They are associated with three families of projectors: $\{P_n\}$, $\{P_{jk}\}$, and $\{P'_{jk}\}$. Additional relations between the probabilities follow from 
\be
x_{jj}= p_j =p_{jj}-p_{j},\quad
y_{jj}
=
0=
p'_{jj}-p_{j}
\ee
and the resulting formula
$
p_j=p'_{jj}=p_{jj}/2.
$
Let us note that $\{P_n\}$ is complete, i.e. $\sum_{n=1}^{\dim \cal H}P_n=\bm 1$, 
$\sum_{n=1}^{\dim \cal H}p_n=1$. In order to understand completness properties of 
$\{P_{jk}\}$ and $\{P'_{jk}\}$ we introduce, for $j<k$, two additional types of vectors and their associated projectors:
\be
|jk^\perp\rangle =\frac{1}{\sqrt{2}}\big(|j\rangle-|k\rangle\big),\quad 
|jk'{}^\perp\rangle =\frac{1}{\sqrt{2}}\big(|j\rangle+i|k\rangle\big), 
\ee
$P_{jk}^\perp=|jk^\perp\rangle\langle jk^\perp|$, 
$P_{jk}'{}\!\!^{\perp}=|jk'{}^\perp\rangle\langle jk'{}^\perp|$.
The completeness relations for $\{P_{jk}\}$ and $\{P'_{jk}\}$ follow from the formula
\be
P_{jk}+P_{jk}^\perp&=& P_j+P_k=P'_{jk}+P_{jk}'{}\!\!^{\perp}.
\ee
The non-Kolmogorovity of the set of probabilities can be illustrated by means of uncertainty relations. Indeed, for any three operators satisfying $[A,B]=iC$ one can prove the uncertainty relation 
$
\Delta A \Delta B\geq \frac{1}{2}|\langle C\rangle|
$
where $\Delta A =\sqrt{\langle A^2\rangle-\langle A\rangle^2}$, $\langle A\rangle=
\Tr (A\rho)$, etc. If $A=P=P^2$ is a one-dimensional projector then $\langle P\rangle=p$ is a probability and one finds $\Delta P=\sqrt{p(1-p)}$. 
We say that two propositions $P_1$ and $P_2$ are complementary if  $\Delta P_1\Delta P_2
\geq \varepsilon>0$. 

In order to show that the propositions, say 
${P}_{j}$ and ${P}_{jk}$, are complementary, we compute 
\be
[{P}_{j},{P}_{jk}]
&=&
\textstyle{\frac{1}{2}}\big(|j\rangle\langle k|
-
|k\rangle\langle j|\big)
=iC
\ee
and 
$\langle C\rangle=y_{kj}$. Finally, 
\be
\sqrt{p_{j}(1-p_{j})}\sqrt{p_{jk}(1-p_{jk})}
\geq \textstyle{\frac{1}{2}}|y_{kj}|.
\ee
The variable $y_{kj}$ measures complementarity of $P_{j}$ and $P_{jk}$.

In supersymmetric projector model we start with $\gamma: {\cal H}\to {\cal H}$ and define the `supercharge' 
$Q:{\cal H}\oplus{\cal H}\to{\cal H}\oplus{\cal H}$,
\be
Q 
=
\left(
\begin{array}{cc}
0 & \gamma\\
\gamma^{\dag} & 0
\end{array}
\right),\label{Q}\quad
Q^2
=
\left(
\begin{array}{cc}
\gamma\gamma^{\dag} & 0\\
0 & \gamma^{\dag}\gamma
\end{array}
\right),\quad
\rho
=
\frac{Q^2}{\Tr(Q^2)}
\ee
The situation where there are two types of super-partner density matrices is encountered in semantic analysis  \cite{AC-LSA}. In semantic contexts the two density matrices correspond to words and sentences, respectively, and it is an intriguing issue if a link to DNA is here accidental. We use the model to describe the process of unwinding of two helices. 

Any of the probabilities we have discussed above is consistent with the frequency interpretation of experiments since always $0\leq p\leq 1$. The subleties related to different models become visible when one comes to joint and conditional probabilities, since only with at least two propositions the issues of non-commutativity become relevant. 

The act of conditioning by a positive answer to the question represented by 
a projector $P$ changes density matrices according to 
\be
\rho
\mapsto 
\rho_1=
\frac{P \rho P}{\Tr (P \rho P)}.
\ee
Joint probability of successes in a sequence of measurements $P_1$, $P_2$, $\dots$, $P_k$ reads
\be
p(P_1\cap \dots \cap P_k)
&=&
\Tr (P_{1}\dots P_{k-1}P_kP_{k-1}\dots P_{1} \rho).\nonumber
\ee
Here $P_1$ is the first measurement and $P_k$ --- the last. Since projectors belonging to different maximal sets do not all commute with one another, the joint probabilities have to be associated with noncommutativity of some sort. In molecular context the most natural noncommutativity is associated with chemical reactions: It is typical that the reactions $X\to Y$ and $Y \to X$ involve different kinetic constants. In particular, it is possible that the chain 
$X\to Y\to Z$ of elementary reactions is possible, whereas the direct reaction $X\to Z$ is not allowed. One can see here an analogy to the well known Malusian experiment with photons and three polarizers. It is worth mentioning that the very fact that some diagrams of the form $X\to Y$ are not allowed may imply a Hilbert space structure \cite{Finkelstein}, of course under some asumptions about the dynamics. 

\section{Soliton kinetic equations}

Let us characterize soliton kinetic equations in a general way. In explicit examples we will work only with a particular case, but since the subject is not widely known we find it important to set our results in a more general context. For further technicalities the readers are referred to \cite{UC,CCU}. 
Consider an operator-valued function 
\be
X_\lambda=\sum_{n=0}^L \lambda^n X_{(n)}+\sum_{n=1}^M \lambda^{-n} X_{(-n)}
\ee
of a complex spectral parameter $\lambda$. 
An invertible operator $D_\lambda$ is a Darboux matrix for $X_\lambda$ if 
\be
X_\lambda[1]
&=&
D_\lambda^{-1}X_\lambda D_\lambda
=\sum_{n=0}^L \lambda^n X_{(n)}[1]+\sum_{n=1}^M \lambda^{-n} X_{(-n)}[1]\label{X[1]}
\ee
where the $X_{(n)}[1]$, $X_{(-n)}[1]$ are $\lambda$-independent
\cite{ZS,NZ,Levi,Mikh,Neu,Cieslinski}. 
In this paper we work with
\be
D_\lambda
&=&
\bm 1+\frac{\nu-\mu}{\mu-\lambda}P\label{D}
\ee
where $P=P^2$ is a projector constructed in terms of right and left eigenvectors of, respectively, $X_\mu$ and $X_\nu$
\cite{LU,U,L,Clifford1,Clifford2}. 
If $\nu=\bar\mu$ and $P=P^{\dag}$ then 
$D_\lambda$ is unitary: $D_\lambda^{-1}=D_\lambda^{\dag}$. Of particular importance is the case 
\be
X_{(\lambda)}=\sum_{n=0}^L \lambda^n X_{n}\label{X[1]+}
\ee
since then putting $\lambda=0$ in (\ref{X[1]+}) we obtain
\be
X_{(0)}[1]=X_0[1]=D_0^{-1}X_0D_0=D_0^{-1}X_{(0)}D_0.
\ee
In later applications to kinetic equations we will work with unitary $D_\lambda$ and 
$X_{0}=\rho$. 

Now consider two operator-valued functions $F(X)$ and $G(X)$ restricted only by 
\be
{[F(X),X]}&=&[G(X),X]=0,\\
\label{fg}
F(TXT^{-1})&=&TF(X)T^{-1},\\
G(TXT^{-1})&=&TG(X)T^{-1},
\ee
for any $T$. The Lax pair is
\be\label{Lax}
z_{\lambda}\psi_\lambda=\psi_\lambda X_{(\lambda)} 
,\quad
-i\dot\psi_\lambda =\psi_\lambda Y_{(\lambda)}
\ee
where $z_{\lambda}$ are complex numbers, and 
\be
\label{H}
X_{(\lambda)} &=&\sum_{k=0}^N\lambda^kX_k,\\
\label{ABC}
Y_{(\lambda)}
&=&\sum_{k=0}^L\lambda^k A_k+\sum_{k=1}^M\frac{1}{\lambda^k}B_k
=Y_{(\lambda)}(X_0,\dots,X_N),\\
\label{B}
A_k
&=&
\frac{1}{(L-k)!}\left.\left(\frac{d^{L-k}}{d\vs^{L-k}}
F\big(\vs^N X_{(\vs^{-1})}\big)\right)\right|_{\vs=0}
\\
\label{C}
B_k
&=&
\frac{1}{(M-k)!}\left.\left(\frac{d^{M-k}}{d\ve^{M-k}}
G\big(X_{(\ve)}\big)\right)\right|_{\ve=0}
\ee
The compatibility condition for the Lax pair consists on $N+1$ equations
\be\label{eode}
i\dot X_m &=&\sum_{k=m+1}^{N}[X_k,A_{m-k}]+
\sum_{k=0}^m[X_k,B_{k-m}],
\quad 0\le m\le N.\label{general}
\ee
For $k<0$ the operators are defined by means of (\ref{B}) and (\ref{C}).
The equation for $m=0$ is the von Neumann equation in question, and those for $m>0$ are additional relations between other operators occuring in the von Neumann equation.

The Darboux matrix defines a Darboux transformation for the Lax pair and its compatibility conditions as follows. 
Take $X_{(n)}[1]$ defined via (\ref{X[1]}), and $A_k[1]$, $B_k[1]$ defined by (\ref{B}), (\ref{C}), but with $X_{(n)}$
replaced by $X_{(n)}[1]$. Then
\be
i\dot X_m[1] &=&\sum_{k=m+1}^{N}[X_k[1],A_{m-k}[1]]+
\sum_{k=0}^m[X_k[1],B_{k-m}[1]],
\quad 0\le m\le N,
\ee
and
\be
z_{\lambda}\psi_\lambda[1]=\psi_\lambda[1] X_{(\lambda)}[1] 
,\quad
-i\dot\psi_\lambda[1] =\psi_\lambda[1] Y_{(\lambda)}[1]
\ee
with $\psi_\lambda[1]=\psi_\lambda D_\lambda$.
Restricting the operators $X_m$ to matrices and defining for $X_0=\rho=\rho^{\dag}>0$ the kinetic variables according to the recipes of Sec.~II we arrive at systems of soliton kinetic equations. 

\section{One-field equations}

We now restrict the explicit examples to `one-field' equations, that is, those 
corresponding to $N=1$, i.e.
\be
X_{(\lambda)} &=& X_0+\lambda X_1=\rho+\lambda H.
\ee
The equations are one-field since only $X_0=\rho$ depends on time. 
One of the two dynamical equations reads here $\dot X_1=0$. Examples of $N>1$ equations can be found in \cite{UC,CCU}. 
An interesing class of nonlinearities occurs for $F(X)=0$, $M=1$, $G(X)=-f(X)$.
Then 
\be
i\dot\rho=[H,f(\rho)].\label{f(rho)}
\ee
The right-hand-side of (\ref{f(rho)}) can be rewritten as $[H_f(\rho),\rho]$, with 
$H_f(\rho)$ whose explicit form depends on $f$, showing that this is indeed a von Neumann equation.
The equation itself was introduced in \cite{MC97} and later discussed in the context of non-extensive statistics in \cite{MCJN99}. The soliton technique of integration based on Darboux-covariant Lax representation was introduced in \cite{SLMC98} and a class of 1-soliton self-switching solutions was there found for $f(\rho)=\rho^2$. 
The Lax-pair representation and Darboux covariance properties of the general form
were established in \cite{UCKL} and explicit self-switching solutions were there found for 
$f(\rho)=\rho^q-2\rho^{q-1}$, where $q$ is an arbitrary real number. 
The explicit form of the Lax pair is simple
\be
i\dot\psi_\lambda =\frac{1}{\lambda}\psi_\lambda f(\rho),\quad
z_{\lambda}\psi_\lambda =\psi_\lambda (\rho+\lambda H).
\ee
In the context of autocatalytic networks the examples of interest involve third order polynomial nonlinearities: $f(\rho)=a_3\rho^3+a_2 \rho^2+a_1\rho$.
Let us choose $f(\rho)=\rho^3-2\rho^2$.
The equation can be written in two equivalent forms
\be
i\dot\rho
&=&
[H,\rho^3-2\rho^2]=[H(\rho),\rho],\\
H(\rho)
&=&
\rho^2 H+\rho H\rho+H\rho^2-2(H\rho+\rho H)
\ee
Taking 
\be
H=\left(
\begin{array}{ccc}
0 & 0 & 0\\
0 & k & 0\\
0 & 0 & 2k
\end{array}
\right)
\ee
and performing the splitting (\ref{p+ip'}) we obtain six coupled kinetic equations with nonlinearities of third order. A closer look at their structure reveals that the reduction of variables from six to four can be obtained if one imposes the constraint
\be
\rho_{23} = z\rho_{12},\quad
\rho_{32} = \bar z\rho_{21},\quad
\rho_{11}=\rho_{33},\quad
|z| = 1.
\ee
Assuming the above reduction with $z=-1$ and introducing the four variables, 
\be
A &=& [A]-\frac{1}{2}p_1-\frac{1}{2}p_2=x_{12}\\
B &=& [B]-p_1=x_{13}\\
C &=& [C]-\frac{1}{2}p_1-\frac{1}{2}p_2=y_{12}\\
D &=& [D]-p_1=y_{13},
\ee
where $x_{ij}$ and $y_{ij}$ are, respectively, real and imaginary parts of $\rho_{ij}$, $[A]$, $[B]$, $[C]$, $[D]$, are nonnegative for all times and $p_1=\rho_{11}$, $p_2=\rho_{22}$ are constants, we can rewrite the von Neumann equation as the set of four kinetic equations
\begin{widetext}
\be
{[\dot A]} &=& -k \Big(2C^3 +2A^2C - AD(2p_1 + p_2-2) 
\nonumber\\
&+& 
          C\big(B^2 +  D^2 - 2 p_1 + p_1^2 - 2p_2 + p_1 p_2 + 
                p_2^2 + B(2 p_1 + p_2-2)\big)\Big)\\
{[\dot B]} &=& 
    -2k\Big(D^3+2A^2 D + D(B^2 + 2C^2  - 4p_1 + 3p_1^2) - 
          2AC(2p_1 + p_2-2)\Big)\\
{[\dot C]} &=& k\Big(2A^3 - CD(2 p_1 + p_2-2) \nonumber\\
&+& 
          A\big(B^2 + 2 C^2 + D^2 - 2 p_1 + p_1^2 - 2p_2 + p_1 p_2 + 
                p_2^2 - B(2 p_1 + p_2-2)\big)\Big)\\
{[\dot D]} &=&
    2k\Big(B^3 + B\big(2C^2 + D^2 + p_1(-4+ 3p_1)\big) 
    \nonumber\\
    &+& 
          A^2(2 + 2 B - 2 p_1 - p_2) + 
          C^2(2 p_1 + p_2-2)\Big)
\ee
\end{widetext}
It is possible to derive this system of kinetic equations from a system of elementary reactions of order not higher than three. 
This system of equations is a particular case, for $q=3$, of the general equation 
$
i\dot\rho
=
[H,\rho^q-2\rho^{q-1}].
$
Its solution for an arbitrary value of $q$, found in \cite{UCKL} by means of Darboux transformations, reads
\be
\rho
&=&
\left(
\begin{array}{ccc}
3/2 & -\xi(t) & \zeta(t)\\
-\bar\xi(t) & 7/4 & \xi(t)\\
\bar\zeta(t) & \bar\xi(t) & 3/2
\end{array}
\right),\label{sol q}\\
\xi(t)
&=&
\frac{(-3i+\sqrt{3})e^{\omega_q (t-t_0)}}
{4\sqrt{2}\big(1+e^{2\omega_q (t-t_0)}\big)}e^{ikt},\\
\zeta(t)
&=&
\frac{1-i\sqrt{3}-2e^{2 \omega_q(t-t_0)}}
{4\big(1+e^{2 \omega_q (t-t_0)}\big)}e^{2ikt},\label{sol q2}
\ee
$\omega_q=[(4/7)^{1-q}-1]k/\sqrt{3}$. This solution is not normalized and hence the kinetic variables have be to rescaled if one is interested in probablities: $p_A=[A]/\Tr\rho$, and so on. Properties of the solution are illustrated by the figures.
The scale of simplification and the efficiency of the von Neumann formalism become evident if one compares the von Neumann and kinetic forms of the same dynamics.

\section{Fluctuations via Darboux transformations}

The set of solutions of von Neumannn kinetic equations is naturally equipped with the  Hilbert-Schmidt norm.
The distance between two solutions $\rho(t)$ and $\tilde\rho(t)$ is
\be
\parallel \rho(t)-\tilde\rho(t)\parallel=\sqrt{\Tr\big(\rho(t)-\tilde\rho(t)\big)^2}.
\ee
The solutions (\ref{sol q}) describe families of distorted helices. We will now explicitly show that for any distorted helix 
(\ref{sol q})--(\ref{sol q2}) there exist  two families of undeformed helices corresponding to two different solutions $\rho_\pm(t)$ of the same equation, such that (i) eigenvalues of $\rho(t)$, $\rho_+(t)$, and $\rho_-(t)$ are identical, (ii) 
$\rho(t)$ and $\rho_\pm(t)$ are linked by Darboux transformations, (iii) $\rho(t)$ and 
$\rho_\pm(t)$ belong to the same symplectic leaf of the Lie-Poisson structure corresponding to the dynamics (i.e. their Poisson-reduced dynamics occurs in the same phase-space), and (iv)
\be
\parallel \rho(t)-\rho_\pm(t)\parallel<\varepsilon
\ee
for any $\varepsilon>0$ and for an infinitely long period of time $t$. In other words, an arbitrarily long part of an undeformed helix is arbitrarily close to a helix that involves a defect. In principle, a small fluctuation of the undeformed helix, at some $t$ and along a symplectic leaf of the dynamics, may map it into a helix with a defect. However, if one restricts the analysis to a finite-length helix $t_1<t<t_2$ (corresponding to a finite number of steps $t_n=n\Delta t$), then the defect may or may not become visible, because different Darboux-transformation fluctuations of the same $\rho(t)$ result in helices with different locations of the defect (parametrized by $t_0$). 
How quickly will the defect occur depends on what kind of a Darboux transformation is responsible for the fluctuation $\rho(t)\mapsto \tilde\rho(t)$.
To see how this works consider
\be
\rho_\pm(t)
=
\left(
\begin{array}{ccc}
3/2 & 0 & \zeta_\pm(t)\\
0 & 7/4 & 0\\
\bar\zeta_\pm(t) & 0 & 3/2
\end{array}
\right),\label{sol q pm}\quad
\zeta_+(t)
=
-\frac{1}{2}
e^{2ikt},\quad
\zeta_-(t)
=
\frac{1-i\sqrt{3}}
{4}e^{2ikt}.
\ee
Density matrices $\rho_\pm(t)$ are the asymptotic forms of $\rho(t)$ for 
$-\omega_qt_0\to\pm\infty$. These two $q$-independent asyptotes also satisfy 
$i\dot\rho_\pm=[H,\rho_\pm^q-2\rho_\pm^{q-1}]$ for any real $q$, which follows from the fact that they can be obtained from the same solution by Darboux transformations differing by initial conditions for the Lax pair. This immediately implies that all these density matrices are isospectral and possess the same values of the Casimirs $\Tr(\rho^n)$. The Casimirs allow to reduce the Lie-Poisson structure to the same symplectic leaf, and thus the fluctuations occur  in the same phase-space. The corresponding Hamiltonian function is given by $\Tr H f(\rho)$. 

\section{Formation of open state in the supersymmetric model}

The supersymmetric model naturally leads to the double-helix structure. 
Consider an arbitrary operator $\gamma$ whose relation to a density matrix is $\rho=\gamma^{\dag}\gamma$. A superpartner of $\rho$ is defined in the usual way as $\tilde\rho=\gamma\gamma^{\dag}$. This type of structue is encountered in Latent Semantic Analysis where $\rho$ and 
$\tilde\rho$ represent word-vector and sentence-vector density matrices \cite{AC-LSA}.  

We know that the Darboux transformation acts on density matrices by $\rho \mapsto \rho_1=
D_0^{\dag}\rho D_0$. This is equivalent to $\gamma\mapsto \gamma_1=\gamma D_0$, 
if $\rho_1=\gamma_1^{\dag}\gamma_1$, or the supercharge transformation
\be
Q 
\mapsto
\left(
\begin{array}{cc}
1 & 0\\
0 & D_0^{\dag}
\end{array}
\right)
\left(
\begin{array}{cc}
0 & \gamma\\
\gamma^{\dag} & 0
\end{array}
\right)
\left(
\begin{array}{cc}
1 & 0\\
0 & D_0
\end{array}
\right)
=
\left(
\begin{array}{cc}
0 & \gamma_1\\
\gamma_1^{\dag} & 0
\end{array}
\right)
=
Q_1, \nonumber
\ee
where $Q$ is given by (\ref{Q}). 
The action of the Darboux transformation on the super-density matrix is
\be
Q_1^2
=
\left(
\begin{array}{cc}
\gamma_1\gamma_1^{\dag} & 0\\
0 & \gamma_1^{\dag}\gamma_1
\end{array}
\right)
=
\left(
\begin{array}{cc}
\gamma\gamma^{\dag} & 0\\
0 & D_0^{\dag}\gamma^{\dag}\gamma D_0
\end{array}
\right)
=
\left(
\begin{array}{cc}
\tilde\rho & 0\\
0 & \rho_1
\end{array}
\right)
\ee
since $D_0$ is unitary. At the level of super-density matrices the Darboux transformation acts by
\be
\left(
\begin{array}{cc}
\tilde\rho & 0\\
0 & \rho
\end{array}
\right)
\mapsto
\left(
\begin{array}{cc}
\tilde\rho & 0\\
0 & \rho_1
\end{array}
\right).
\ee
Of particular interest is the symmetric case $\gamma\gamma^{\dag}=\gamma^{\dag}\gamma$. This is trivially satisfied if $\gamma=\gamma^{\dag}$ and $\gamma$ itself is a supercharge for $\rho=\tilde\rho$. Such a $\gamma$ always exists since $\rho$, being a positive operator, possesses a unique square root 
$\sqrt{\rho}=\gamma$. The Darboux transformation breakes this symmetry as follows
\be
\left(
\begin{array}{cc}
\rho & 0\\
0 & \rho
\end{array}
\right)
\mapsto
\left(
\begin{array}{cc}
\rho & 0\\
0 & \rho_1
\end{array}
\right).
\ee
It is interesting that any sequence of Darboux transformations is here necessarily cyclic in the sense that 
\be
\left(
\begin{array}{cc}
\rho & 0\\
0 & \rho
\end{array}
\right)
\to
\left(
\begin{array}{cc}
\rho & 0\\
0 & \rho_1
\end{array}
\right)
\to
\dots
\to
\left(
\begin{array}{cc}
\rho & 0\\
0 & \rho_n
\end{array}
\right)
\to
\left(
\begin{array}{cc}
\rho & 0\\
0 & \rho
\end{array}
\right)\nonumber
\ee
with $n$ being related to the dimension of the matrix $\rho$. To understand why this has to happen let us recall that the Darboux transformation involves two solutions  of a Lax pair: The one that is transformed, and the one that is used to define the transformation. The transformation is trivial if the two solutions are linearly dependent. It is clear that the number of linearly independent solutions is limited by the dimension of the linear space they live in. In the context of $3\times 3$ solutions we find 
\be
\rho\to\rho_1\to\rho_2\to\rho_3\to\rho,\label{scheme}
\ee
no matter what (nontrivial) transformations have been performed, and two of these four density matrices have the asymptotic form $\rho_\pm(t)$. This is why after four Darboux fluctuations the system returns to its original state. 

For super-density matrices this process is strikingly similar to what happens with DNA during formation of an open state. Initially we have two identical strands described by the symmetric state 
$\rho=\tilde\rho$. The first nontrivial fluctuation maps it into the pair $(\rho,\rho_1)$ whose evolution looks as either Fig.~1 or Fig.~2 where $\rho$ and $\rho_1$ correspond to the two helices. Let us note, that the helices are practically overlapping over arbitrarily long parts, and therefore an arbitrarily small fluctuation may turn one into another. If the fluctuation is of the form shown in Fig.~1 then somewhere in the future the two helices will unwind, unless a new fluctuation will prevent it. If the fluctuation is of the form shown in Fig.~2, then the helix switches into a helix that wants to return to the asymptote, and we will not see a change of form. However, if the helices disentangled, then the helix approaches the other asymptote, and a subsequent small fluctuation will not be observed. Finally, another fluctuation maps it into a state that starts to return asymptotically to the original state. 

The fluctuations deform the helices according to the scheme (\ref{scheme}). However, we do not know which Darboux transformation is responsible for $\rho\to\rho_1$, say. The point is that this is not really important! Fig.~5 shows two helices corresponding to two randomly chosen Darboux transformations. The random element is {\it when\/} do the helices start to unwind. All these helices have the same asymptotic forms and 
the part where the two helices look different is the one where the Hilbert-Schmidt distance is maximal. 

One can illustrate this phenomenon if one plots the Hilbert-Schmidt distance between two helices whose forms differ by the parameter $t_0$. Fig.~6 shows the plot of the Hilbert-Schmidt distance between two solutions that were produced from the same seed solution by two different Darboux transformations, as a function of $t$ and the fluctuation parameter $t_0$ (in one solution $t_0=0$ and is arbitrary in the other one). Two such density matrices effectively differ from each other in a finite time interval. 

\section{Summary and perspectives}

We have started with the observation that a system evolving into helical structures has to resemble an oscillator, at least at a sufficiently abstract level of description. Armed with this intuition and the fact that dynamical equations of DNA must be kinetic in some sense we have arrived at equations of a von Neumann type. Then, imposing soliton-type integrability, we restricted the class to Darboux-covariant von Neumann equations. 

The formalism we have proposed has many possible continuations. We have touched only the tip of the iceberg. Even the one-field equations are more general than those with the right-hand side $[H,f(\rho)]$. 
As shown in \cite{CCU}, the class contains also Toda and other integrable lattices as particular cases. 
Solutions can be found by the same Darboux transformation, but one has to employ a different explicit technique of finding nontrivial seed solutions (cf. \cite{KCL}). 
Still another technique of finding appropriate seed solutions that generalizes to {\it arbitrary\/}  
$H$ and $f(\rho)$ the method we have worked with, was introduced in \cite{Kuna}. When applied to explicit matrix examples the method leads again to qualitatively the same solutions as those we already know. One can risk the statement that the type of behavior 
(unwinding and switching between a finite number of states) we have described is {\it generic\/} for the whole class of integrable kinetic equations. In the context of genericity one should mention also the stability of solutions for $f(\rho)=\rho^q-2\rho^{q-1}$ under modifications of $q$ (cf. the discussion in \cite{UCKL}). This means that the behaviour of the helix is basically the same for $q=3$ and, say, $q=1000$, or a non-integer $q$. If one tried to write down kinetic equations corresponding to $q=1000$ one would have found the set of equations with polynomials of order $1000$ and thousands of terms. Analogously, there would have been thousands of elementary very-high order reactions that, effectively lead to almost identical replication and unwinding as $q=3$ or $q=2$. This fact may explain why incredibely simplified kinetic schemes one employs in replication retworks lead to evolutions that reasonably approximate various effects one expects in DNA-type systems. 

\acknowledgments
The work of MC is a part of the Polish Ministry of Scientific Research and Information Technology (solicited) project PZB-MIN 008/P03/2003. We acknowledge the support  of the Flemish Fund for Scientific Research (FWO Project No. G.0335.02).
MC is grateful to professor Stanis{\l}aw Koter for his comments on chemical harmonic oscillations.

\begin{figure}\label{helix-minus}
\includegraphics{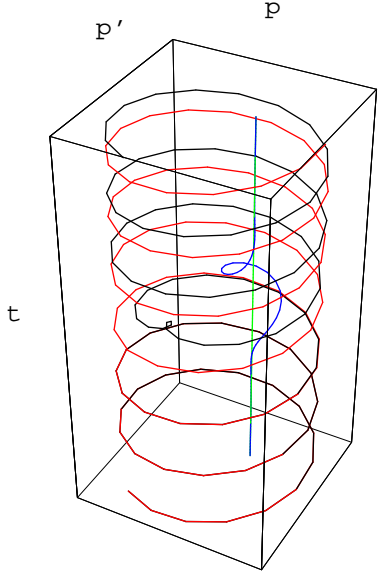}
\caption{Formation of open state. The red, black, blue, and green helices represent, respectively, 
the matrix elements $\rho_-(t)_{13}$,
$\rho(t)_{13}$, $\rho_-(t)_{12}$, and
$\rho(t)_{12}$. $\rho(t)$ and $\rho_-(t)$ are given by (\ref{sol q}) and (\ref{sol q pm}).
The helices practically overlap in the past and unwind in a neighbourhood of $t=0$. The four helices behave as if the external ones were the tapes, and the internal ones represented the read/write head.} 
\end{figure}
\begin{figure}\label{helix-plus}
\includegraphics{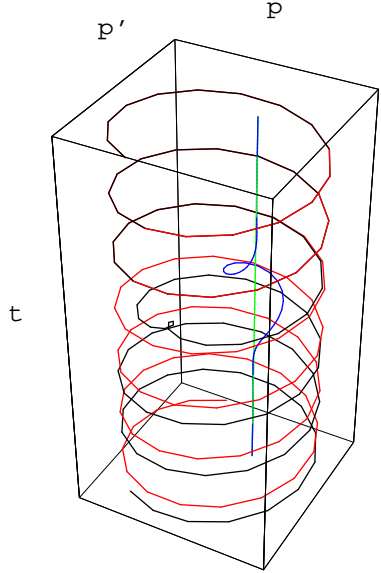}
\caption{Formation of closed state. The red, black, blue, and green helices represent, respectively, 
the matrix elements $\rho_+(t)_{13}$,
$\rho(t)_{13}$, $\rho_+(t)_{12}$, and
$\rho(t)_{12}$. $\rho(t)$ and $\rho_+(t)$ are given by (\ref{sol q}) and (\ref{sol q pm}). The helices practically overlap in the future and get wound up in a neighbourhood of $t=0$.} 
\end{figure}
\begin{figure}\label{rzut-t}
\includegraphics{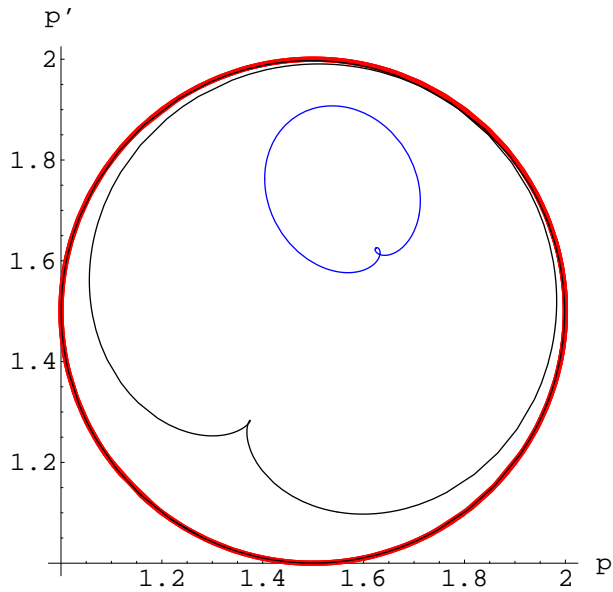}
\caption{Projection of Fig.~1 and Fig.~2 on the $(p,p')$-plane.} 
\end{figure}

\begin{figure}\label{H-S-dist}
\includegraphics{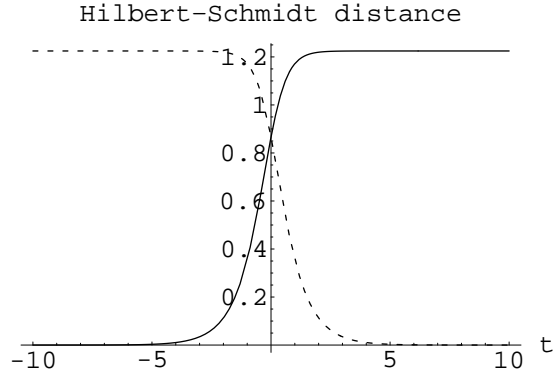}
\caption{Hilbert-Schmidt distances $\parallel\rho(t)-\rho_+(t)\parallel$ (dotted), and 
$\parallel\rho(t)-\rho_-(t)\parallel$ (full) as functions of time.
$\rho(t)$ and $\rho_\pm(t)$ are given by (\ref{sol q}) and (\ref{sol q pm}).} 
\end{figure}
\begin{figure}\label{helix-dwie}
\includegraphics{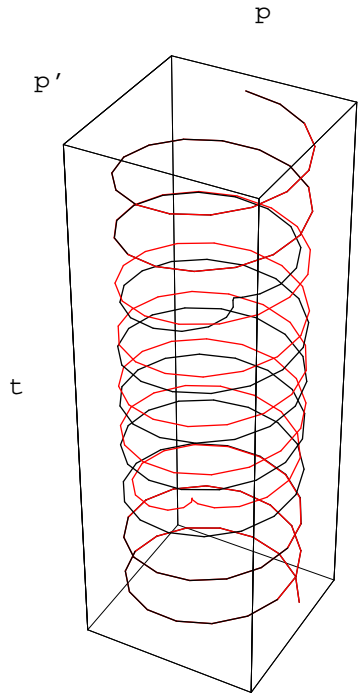}
\caption{Two helices obtained from the same $\rho$ by means of two nontrivial Darboux transformations. The distance between two such helices is shown in the next figure.} 
\end{figure}
\begin{figure}\label{H-S-dist-t-t0}
\includegraphics{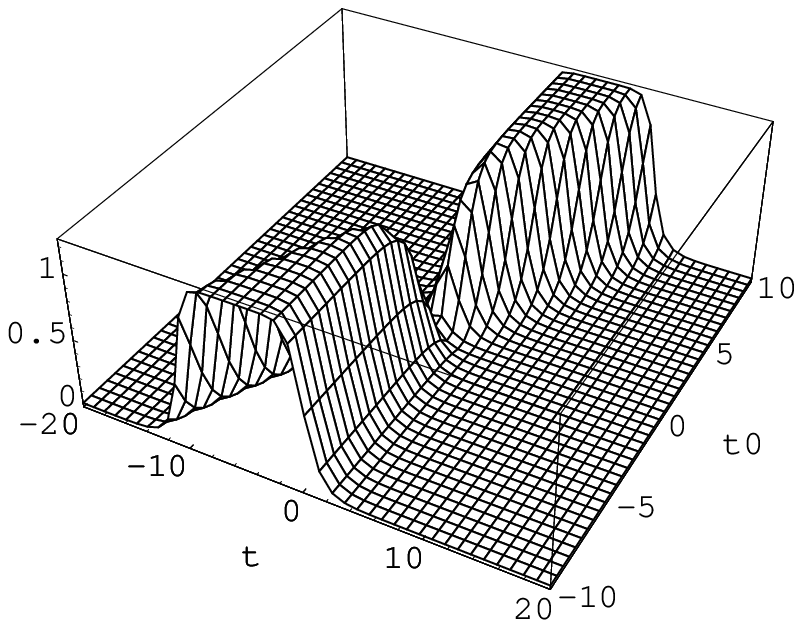}
\caption{Hilbert-Schmidt distance between two solutions corresponding, respectively, to 
$t_0=0$ and an arbitrary $t_0$, plotted as a function of time $t$ and the parameter $t_0$. Solutions differing by different values of $t_0$ are obtained from the same seed solution by two different Darboux transformations. The distance is maximal in the part of the plot where the two corresponding helices are unwound (the `open state').} 
\end{figure}


\begin{thebibliography}{99}
\bibitem{BL}C. H. Bennett, R. Landauer, Scientific American {\bf 253}, 48-56 (1985).
\bibitem{Adleman}L. M. Adleman, Science {\bf 266}, 1021 (1994).
\bibitem{Englanger}S. W. Englander {\it et al.\/}, Proc. Nat. Acad, Sci. 
{\bf 77}, 7222 (1980).
\bibitem{Yak}L. V. Yakushevich, {\it Nonlinear Physics of DNA\/} (Wiley, New York, 1998).
\bibitem{Yak1}L. V. Yakushevich, J. Biosci. {\bf 26}, 305 (2001).
\bibitem{Hartmann}B. Hartmann, W. J. Zakrzewski, J. Nonlin. Math. Phys. {\bf 12} (2005) --- in print.
\bibitem{Eigen}M. Eigen, Natuwissenschaften {\bf 58}, 465 (1971).
\bibitem{Schuster}P. Schuster and K. Sigmund, J. Theor. Biol. {\bf 100}, 533 (1983).
\bibitem{HS}J. Hofbauer and K. Sigmund, {\it Dynamical Systems and the Theory of Evolution\/} (Cambridge University Press, Cambridge, 1988).
\bibitem{SQ1}K. Steiglitz, I. Kamal, and A. H. Watson, IEEE Transactions on Computers {\bf 37}, 138 (1988).
\bibitem{SQ2}M. H. Jakubowski, K. Steiglitz, and R. K. Squier, Complex Systems {\bf 10}, 1 
(1996).
\bibitem{SQ3}M. H. Jakubowski, K. Steiglitz, and R. K. Squier, Phys. Rev. E {\bf 58}, 6752 
(1998); Phys. Rev. E {\bf 56}, 7267 (1997)
\bibitem{Symes}W. Symes, Physica D {\bf 4}, 275 (1982).
\bibitem{Przybylska1}M. Przybylska, Linear Algebra Appl. {\bf 346}, 155 (2002).
\bibitem{Przybylska2}M. Przybylska, Future Generation Comp. Syst. {\bf 19}, 1165 (2003).
\bibitem{Salle} V.B. Matveev, M.A. Salle, {\it Darboux Transformations 
and Solitons\/} (Springer--Verlag, Berlin--Heidelberg, 1991).
\bibitem{SKoter}S. Koter, private communication.
\bibitem{PRE03}D. Aerts, M. Czachor, L. Gabora, M. Kuna, A. Posiewnik, J. Pykacz, and M. Syty, Phys. Rev. E {\bf 67}, 051926 (2003).
\bibitem{Pryk}V. V. Gafiychuk and A. K. Prykarpatsky, J. Nonlin. Math. Phys. {\bf 11}, 
350 (2004).
\bibitem{AC-LSA}D. Aerts and M. Czachor, J. Phys. A: Math. Gen. {\bf 37}, L123 (2004).
\bibitem{Finkelstein}D. R. Finkelstein, {\it Quantum Relativity: A Synthesis of the Ideas of Einstein and Heisenberg\/} 
(Springer, Berlin, 1996).
\bibitem{UC}N. V. Ustinov and M. Czachor, in {\it Probing the Structure of Quantum Mechanics: Nonlinearity, Nonlocality, Computation, and Axiomatics\/}, Eds. D. Aerts, M. Czachor, and T. Durt (World Scientific, Singapore, 2002), 335.
\bibitem{CCU}J. L. Cie\'sli\'nski, M. Czachor, and  N. V. Ustinov, J. Math. Phys. {\bf 44}, 1763 (2003).
\bibitem{ZS}V. E. Zakharov and A. B. Shabat, Funk. Anal. Pril. {\bf 13}, 13 (1979) (in Russian).
\bibitem{NZ}S.P.\,Novikov, S.V.\,Manakov, L.P.\,Pitaevski and V.E.\,Zakharov,
{\it Theory of Solitons, the Inverse Scattering Method\/}
(Consultants Bureau, New York, 1984).
\bibitem{Levi}D.\,Levi, O.\,Ragnisco, and M.\,Bruschi, Nuovo Cim. {\bf 58A}, 
56 (1980).
\bibitem{Mikh}A.V.\,Mikhailov, Physica {\bf D\,3}, 73 (1981).
\bibitem{Neu}G.\,Neugebauer and D.\,Kramer, J. Phys. A: Math. Gen. 
{\bf 16}, 1927 (1983).
\bibitem{Cieslinski}J. L.\,Cie\'sli\'nski, J. Math. Phys. {\bf 32}, 2395 (1991);
ibid. {\bf 36}, 5670 (1995).
\bibitem{LU}S.B.\,Leble and N.V.\,Ustinov, Solitons of nonlinear equations
associated with degenerate spectral problem of the third order,
in: Nonlinear Theory and its Applications, Eds. M.\,Tanaka and T.\,Saito
(World Scientific, Singapore, 1993) v.2, pp.547-550.
\bibitem{U}N. V. Ustinov, J. Math. Phys. {\bf 39}, 976 (1998).
\bibitem{L}S. B. Leble, Computers Math. Applic. {\bf 35}, 73 (1998).
\bibitem{Clifford1}J. L. Cie\'sli\'nski, J. Phys. A: Math. Gen. {\bf 33}, L363 (2000).
\bibitem{Clifford2}W. Biernacki and J. L. Cie\'sli\'nski, Phys. Lett. A {\bf 288}, 167 (2001).
\bibitem{MC97}M.\,Czachor, Phys. Lett. {\bf 225A}, 1 (1997).
\bibitem{MCJN99}M.\,Czachor and J.\,Naudts, Phys. Rev. {\bf 59E}, R2497 (1999).
\bibitem{SLMC98}S.B.\,Leble and M.\,Czachor, Phys. Rev. {\bf 58E}, 7091 (1998).
\bibitem{UCKL}N.V.\,Ustinov, M. Czachor, M. Kuna, and S.B.\,Leble, 
Phys. Lett. A {\bf 279}, 333 (2001). 
\bibitem{Math}The solutions were checked by means of Mathematica 4.2.
\bibitem{KCL}M.\,Kuna, M.\,Czachor and S.B.\,Leble, Phys. Lett. {\bf 255A},
42 (1999).
\bibitem{Kuna}M. Kuna, quant-ph/0408048.
\end{thebibliography}
\end{document}